\documentclass[aip,apl,reprint,floatfix,groupedaddress]{revtex4-1}

\usepackage{amsmath}
\usepackage{graphicx}
\usepackage{bm}

\renewcommand{\vector}[1]{\bm{#1}}
\newcommand{\lhm}{LHM}

\begin{document}

\title{Miniaturization of photonic waveguides by the use of left-handed materials}

\author{Philippe Tassin}
\affiliation{Department of Applied Physics and Photonics, %
             Vrije Universiteit Brussel,\\ %
             Pleinlaan 2, B-1050 Brussel, Belgium}

\author{Xavier Sahyoun}
\affiliation{Department of Applied Physics and Photonics, %
             Vrije Universiteit Brussel,\\ %
             Pleinlaan 2, B-1050 Brussel, Belgium}

\author{Irina Veretennicoff}
\affiliation{Department of Applied Physics and Photonics, %
             Vrije Universiteit Brussel,\\ %
             Pleinlaan 2, B-1050 Brussel, Belgium}

\date{January 15, 2008}

\begin{abstract}
We propose the use of a left-handed material in an optical waveguide structure to reduce its thickness well below the wavelength of light. We demonstrate that a layer of left-handed material, added to the cladding of a planar waveguide rather than to its core, allows for good light confinement in a subwavelength thin waveguide. We attribute the observed behavior to the change in phase evolution of electromagnetic waves in the guide. This technique can be used for the miniaturization of photonic integrated circuits.
\end{abstract}

\maketitle

A lot of progress has been made in the field of photonic integrated circuits during the last decades.\cite{Nagarajan-2007} The integration of lasers, modulators and gratings has led to  photonic integrated circuits consisting of several dozens of components, which provide a possibility to address the increasing need for higher data capacity and more functionality in communication systems. Nevertheless, in comparison with their electronic counterparts, their integration density remains low, mainly due to a wavelength condition on the size of their constituents. For example, photonic flip-flops based on microring lasers have a circumference of at least one wavelength to ensure resonance,\cite{Hill-2004} and optical wave\-guides must have their core diameter larger than the wavelength for adequate light confinement.

Recently, the subwavelength miniaturization of photonic components has been addressed by the use of techniques from the fields of plasmonics\cite{Hill-2007} and metamaterials. For instance, it was shown that the classical limitation on the width of a Fabry-Perot resonator can be overcome by the use of left-handed materials (\lhm{}s).\cite{Engheta-2002} These metamaterials with negative permittivity and permeability were experimentally demonstrated in 2001.\cite{Shelby-2001} Since, researchers have improved and scaled down these structures to fabricate \lhm{}s at optical wavelengths,\cite{Dolling-2006,Shalaev-2007} and have proposed metamaterials for various applications, such as electromagnetic cloaking,\cite{Pendry-2006} improved lenses,\cite{Pendry-2000,Tassin-2006} sub-diffraction-limited light beams,\cite{Kockaert-2006,Tassin-2006b} etc. Wave\-guides with only \lhm{} in the core have been studied by Shadrivov \emph{et al.},\cite{Shadrivov-2003} who have shown the existence of surface waves and vortex-like structures in the energy flux. Furthermore, waveguides of thickness of about $\lambda/3$ have been made possible by the use of photonic crystals.\cite{Joannopoulos-1995} Nevertheless, the size of these structures remains limited by diffraction.

When a layer of \lhm{} is inserted into a Fabry-Perot cavity, photons will pass through the right-handed (positive index of refraction) and the left-handed material (negative index of refraction) during each roundtrip, and the phase of an optical wave will therefore evolve forward and backward.\cite{Engheta-2002} The resulting phase compensation allows for a resonance in the cavity with vanishing optical path length, for which the cavity's width can become subwavelength thin. A similar technique has been applied to shielded microwave waveguides with perfect electric conducting boundary conditions.\cite{Engheta-2006}

\begin{figure}[!thb]
\begin{center}
\includegraphics{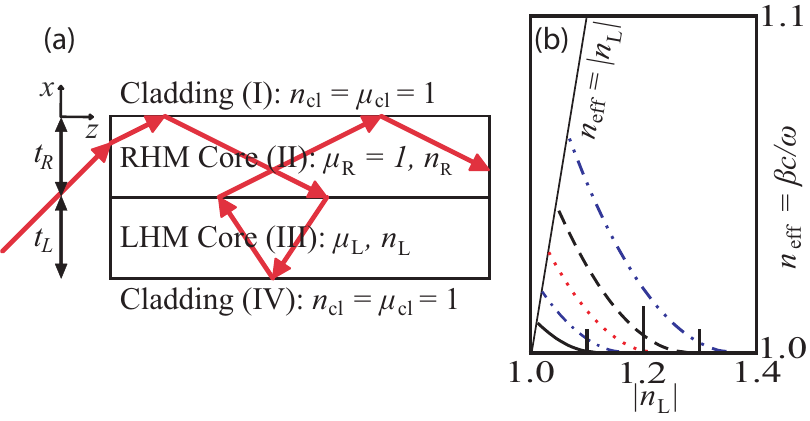}
\caption{(a) Optical waveguide with \lhm{} added to the core. (b) Effective mode index $n_{\mathrm{eff}}=\beta c / \omega$ as a function of index of refraction of the \lhm{} for several values of $t_\mathrm{R}$. Parameters are $n_\mathrm{R} = 1.5$, $t_\mathrm{L} = 0.033\lambda$, $\mu_\mathrm{L} = -0.2$, and $n_\mathrm{cl} = 1.0$.}
\label{Fig:LHMCoreStructure}
\end{center}
\end{figure}

In this Letter, we provide a proof-of-principle of a subwavelength thin optical waveguide---an essential component in integrated circuits---with a mode well confined to its core. In order to investigate if the addition of a \lhm{} to the core could also be used for the miniaturization of photonic waveguides with open boundary conditions, we have studied the planar model of Fig.~\ref{Fig:LHMCoreStructure}(a). The core of this waveguide consists of two layers, one layer contains a right-handed material (RHM), the other layer a left-handed (LHM) one, and is surrounded by identical cladding media at both sides. The thickness of the RHM and \lhm{} layer is, respectively, $t_\mathrm{R}$ and $t_\mathrm{L}$. Due to negative refraction, the rays will trace paths similar to the one plotted in Fig.~\ref{Fig:LHMCoreStructure}(a) (Reflections at the intracore interface are neglected for now, but will be included in the full analysis below.) Within every period, the ray passes through an amount of left-handed and right-handed material. If the difference in optical path length in both materials becomes small enough, one could expect to have a guided mode with subwavelength diameter analogous to the mode in the shielded microwave waveguide.

We have derived analytical guided mode solutions of Maxwell's equations for this geometry. We searched for TE polarized monochromatic waves with an electric field of the form $\vector{E} = \vector{e}_y \zeta\left( x \right) \exp{\left[ i\left( \beta z - \omega t \right) \right]}$, where $\omega$ is the angular frequency and $\beta$ is the propagation constant or the parallel component of the wave vector, which remains conserved throughout all interfaces by virtue of the boundary conditions. Since we consider monochromatic waves, the only effect of the typically strong dispersion observed for \lhm{}s is that the material parameters have to be taken at their values corresponding to the frequency used; dispersion will nevertheless play its role when considering non-monochromatic signals in the waveguide. The mode profile $\zeta\left( x \right)$ in the four layers can be determined from the wave equation:
\begin{eqnarray}
\zeta_\mathrm{I}\left( x \right) &=& A \exp\left( -\kappa_\mathrm{cl} x \right), \label{Eq:Core:WaveI}\\
\zeta_\mathrm{II}\left( x \right) &=& B \cos \left( k_\mathrm{R} x \right) + C \sin \left( k_\mathrm{R} x \right), \label{Eq:Core:WaveII}\\
\zeta_\mathrm{III}\left( x \right) &=& D \cos \left[ k_\mathrm{L} \left( x+t_\mathrm{R} \right) \right] + E \sin \left[ k_\mathrm{L} \left( x+t_\mathrm{R} \right) \right], \label{Eq:Core:WaveIII}\\
\zeta_\mathrm{IV}\left( x \right) &=& F \exp\left[ \kappa_\mathrm{cl} \left( x + t_\mathrm{R} + t_\mathrm{L} \right) \right], \label{Eq:Core:WaveIV}
\end{eqnarray}
where $k_{R} = \sqrt{\omega^2 n_\mathrm{R}^2/c^2-\beta^2}$, $k_{L} = -\sqrt{\omega^2 n_\mathrm{L}^2/c^2-\beta^2}$ and $\kappa_\mathrm{cl} = \sqrt{\beta^2-\omega^2 n_\mathrm{cl}^2/c^2}$. The expressions~(\ref{Eq:Core:WaveI})-(\ref{Eq:Core:WaveIV}), representing waves propagating in the core layers and with exponential tails in the cladding, are valid if $\omega^2 n_\mathrm{cl}^2/c^2 < \beta^2 < \omega^2 n_\mathrm{R,L}^2/c^2$. The amplitudes $A$, $B$, $C$, $D$, $E$ and $F$ can be determined from continuity equations at the boundary layers, although care should be taken when applying these boundary conditions to an interface of the \lhm{} layer, because the time derivative of the electric field changes sign at such an interface. Subsequent elimination of these amplitudes leads to the following dispersion relation:
\begin{equation}
\begin{split}
\frac{\tau_\mathrm{L}}{\tau_\mathrm{R}} \left( \cos k_\mathrm{R} t_\mathrm{R} + \frac{\tau_\mathrm{cl}}{\tau_\mathrm{R}} \sin k_\mathrm{R} t_\mathrm{R} \right) \left( \cos k_\mathrm{L} t_\mathrm{L} - \frac{\tau_\mathrm{L}}{\tau_\mathrm{cl}} \sin k_\mathrm{L} t_\mathrm{L} \right) \\
= \left( \sin k_\mathrm{R} t_\mathrm{R} - \frac{\tau_\mathrm{cl}}{\tau_\mathrm{R}} \cos k_\mathrm{R} t_\mathrm{R} \right) \left( \sin k_\mathrm{L} t_\mathrm{L} + \frac{\tau_\mathrm{L}}{\tau_\mathrm{cl}} \cos k_\mathrm{L} t_\mathrm{L} \right),
\end{split} \label{Eq:Core:DispersionRelation}
\end{equation}
where $\tau_\mathrm{R} = k_\mathrm{R}/\mu_\mathrm{R}$, $\tau_\mathrm{L} = k_\mathrm{L}/\mu_\mathrm{L}$, and $\tau_\mathrm{cl} = \kappa_\mathrm{cl}/\mu_\mathrm{cl}$.

If the impedances and optical path lengths of both core layers are matched (i.e., if $\tau_\mathrm{R} = \tau_\mathrm{L}$ and $k_\mathrm{R} t_\mathrm{R} = - k_\mathrm{L} t_\mathrm{L}$), it can be verified that Eq.~(\ref{Eq:Core:DispersionRelation}) has no solutions. This means that the waveguide with compensated optical path length exhibits no guided modes at all. This must be compared with the all-angle bandgap that is observed for a photonic crystal with zero average refractive index;\cite{Li-2003,Daninthe-2006} the latter condition is indeed equivalent to zero optical path length. Only if $\kappa_\mathrm{cl}\rightarrow \infty$, as applicable for the microwave waveguide, a singularity arises in Eq.~(\ref{Eq:Core:DispersionRelation}) and an infinite number of guided mode solutions appear. From a complete numerical search for solutions to Eq.~(\ref{Eq:Core:DispersionRelation}) performed for different geometrical ($t_{\mathrm{R},\mathrm{L}}$) and material ($n_{\mathrm{R},\mathrm{L}}$, $\mu_{\mathrm{R},\mathrm{L}}$) parameters, we have found that this situation remains for small deviations from the conditions of impedance matching and optical path compensation. However, as soon as either of both conditions is relaxed sufficiently, guided waves can be found, but in this case the effective mode index is always very small [see also Fig.~\ref{Fig:LHMCoreStructure}(b)]. This indicates that it is impossible with the geometry of Fig.~\ref{Fig:LHMCoreStructure}(a) to have a guided mode in a subwavelength thin waveguide with good light confinement. Waveguides with exclusively \lhm{} in the core suffer from the same problem, as is evident from the dispersion relations given by Shadrivov~\emph{et al.}\cite{Shadrivov-2003}

We can attribute this breakdown of guided modes to the additional phase shifts that an optical ray undergoes when reflected at the boundaries of the core. If reflections on the intracore boundary are neglected, it is possible to write every guided mode as the interference pattern of oblique plane waves, or rays. Only those plane waves that interfere constructively lead to a guided wave solution. The phase of a ray changes in one period of the mode by:
\begin{equation}
\Delta\varphi = \frac{2\omega}{c} n_\mathrm{R} t_\mathrm{R} \cos\theta_\mathrm{R} + \frac{2\omega}{c} n_\mathrm{L} t_\mathrm{L} \cos\theta_\mathrm{L} + \Delta\varphi_\mathrm{R} + \Delta\varphi_\mathrm{L}.\label{Eq:Core:ModeCondition}
\end{equation}
In the shielded waveguide, the phase shifts $\Delta\varphi_\mathrm{R}$ and $\Delta\varphi_\mathrm{L}$ associated with reflection on the edges of the structure are zero. But in a photonic waveguide with open boundaries, the rays are confined by total internal reflection, for which the rays do undergo additional phase shifts. In this case, $\Delta\varphi_\mathrm{R}$ and $\Delta\varphi_\mathrm{L}$ can be estimated from Fresnel's formulas.\cite{Azzam-2004} This will not be repeated here; it suffices to recall that the phase shifts are bounded by $-\pi<\Delta\varphi_\mathrm{R,L}<0$. Note that $\Delta\varphi_\mathrm{R}$ and $\Delta\varphi_\mathrm{L}$ have the same sign irrespective of the handedness of the material in which the ray propagates. We can now understand what is happening when we try to compensate the optical path length: the first two terms of Eq.~(\ref{Eq:Core:ModeCondition}) will cancel each other out, but the two reflection phase shifts have the same sign and will add up to a number not satisfying the mode condition $\Delta\varphi = 2\pi m$.

\begin{figure}[!thb]
\begin{center}
\includegraphics{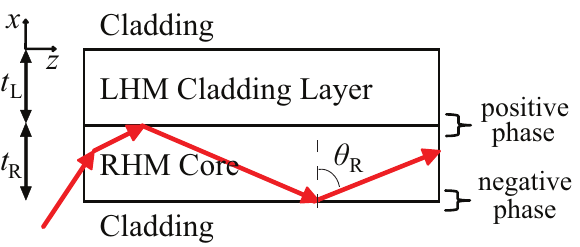}
\caption{Optical waveguide with \lhm{} added to the cladding.}
\label{Fig:WaveguideLHMCladding}
\end{center}
\end{figure}

\begin{figure}[!thb]
\begin{center}
\includegraphics{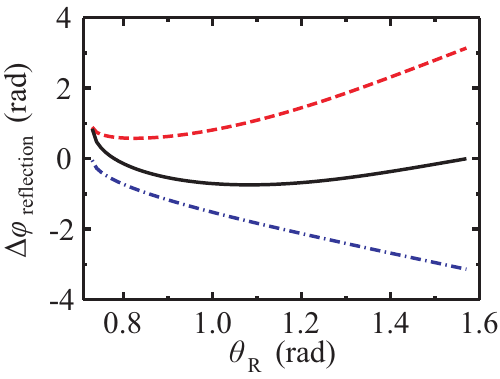}
\caption{Phase shifts due to reflection on the right-handed clad\-ding layer (blue, dash-dotted curve), on the left-handed cladding layer (red, dashed curve), and the total phase shift (black curve) as a function of angle of reflection. Phase compensation is here achieved with a \lhm{} slab of $0.033\lambda$.}
\label{Fig:PhaseShifts}
\end{center}
\end{figure}

We can now conclude that, in order to create a well-confined guided mode in a subwavelength thin photonic waveguide, it is necessary to engineer the phase shift due to reflection on one of its core-cladding interfaces, so as to have a cancellation of the total reflection phase. A careful analysis of Fresnel's formulas has provided us with a solution: one has to add a \lhm{} to the cladding rather than to the core, as in Fig.~\ref{Fig:WaveguideLHMCladding}. An optical ray propagating in the single core layer of this structure will undergo phase shifts due to propagation within the core layer and due to the reflections on the left-handed ($\Delta\varphi_\mathrm{R}$) and right-handed cladding layers ($\Delta\varphi_\mathrm{L}$):
\begin{equation}
\Delta\varphi = \frac{2\omega}{c} n_\mathrm{R} t_\mathrm{R} \cos\theta_\mathrm{R} + \Delta\varphi_\mathrm{L} + \Delta\varphi_\mathrm{R}.\label{Eq:Cladding:ModeCondition}
\end{equation}
For a waveguide with subwavelength thickness, the first term of Eq.~(\ref{Eq:Cladding:ModeCondition}) is negligible. $\Delta\varphi_\mathrm{R}$ is still given by Fresnel's formula and is thus negative. The phase shift $\Delta\varphi_\mathrm{L}$ can be estimated by considering the total internal reflection of an obliquely incident plane wave on the stratified Core-LHM-Cladding structure, which can be regarded as a Fabry-Perot resonator for which the internal and transmitted waves are evanescent. This analysis leads to the following formula for $\Delta\varphi_\mathrm{L}$:
\begin{equation}
\Delta\varphi_{L} = -2 \arctan\left( \frac{\tau_\mathrm{cl}}{\tau_\mathrm{R}} \frac{1+\frac{\tau_\mathrm{L}}{\tau_\mathrm{cl}}\tanh\kappa_\mathrm{L} t_\mathrm{L}}{1+\frac{\tau_\mathrm{cl}}{\tau_\mathrm{L}}\tanh\kappa_\mathrm{L} t_\mathrm{L}} \right),
\end{equation}
where $\kappa_\mathrm{L} = \sqrt{\beta^2 - \omega^2 n_\mathrm{L}^2/c^2}$, $\beta = \omega n_\mathrm{R} \sin\theta_\mathrm{R}/c$, and $\tau_\mathrm{L} = \kappa_\mathrm{L}/\mu_\mathrm{L}$. $\tau_\mathrm{R}$ and $\tau_\mathrm{cl}$ are as defined above. The sign of this phase shift strongly depends on the values of the geometrical and material parameters of the waveguide. Nevertheless, our calculations presented in Fig.~\ref{Fig:PhaseShifts} show that it is indeed possible to create a positive reflection phase shift with a subwavelength thin ($t_\mathrm{L} = 0.033\lambda)$ \lhm{} layer in the cladding.

\begin{figure}[!thb]
\begin{center}
\includegraphics[clip]{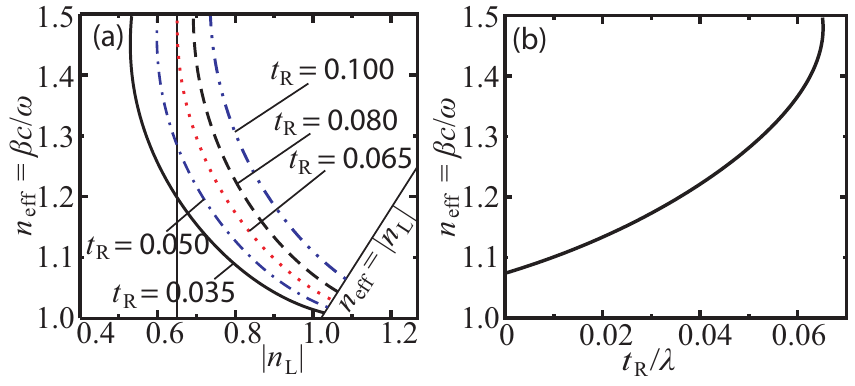}
\caption{Dispersion relation in a waveguide with total thickness of less than $10\%$ of the wavelength. Parameters are the same as in Fig.~\ref{Fig:LHMCoreStructure} and $n_\mathrm{L} = \sqrt{0.4}$.}
\label{Fig:EffectiveModeIndex}
\end{center}
\end{figure}

In order to validate the previous predictions, we have solved Maxwell's equations for waves propagating in the core layer and with exponential tails in the cladding layers of the structure in Fig.~\ref{Fig:WaveguideLHMCladding}, i.e., in the region $\left|n_\mathrm{L,cl}\right| < n_\mathrm{eff} = \beta/(\omega/c) < n_\mathrm{R}$ . This leads to the following dispersion relation determining the propagation constant $\beta$:
\begin{equation}
\begin{split}
&\frac{\tau_\mathrm{cl} + \tau_\mathrm{L}}{\tau_\mathrm{cl} - \tau_\mathrm{L}} \left[ 1 + \frac{\tau_\mathrm{L}}{\tau_\mathrm{R}} \left( \frac{\tau_\mathrm{cl}\sin k_\mathrm{R} t_\mathrm{R} + \tau_\mathrm{R} \cos k_\mathrm{R} t_\mathrm{R}}{\tau_\mathrm{cl}\cos k_\mathrm{R} t_\mathrm{R} - \tau_\mathrm{R} \sin k_\mathrm{R} t_\mathrm{R}} \right) \right] \\
= &\left[ 1 - \frac{\tau_\mathrm{L}}{\tau_\mathrm{R}} \left( \frac{\tau_\mathrm{cl}\sin k_\mathrm{R} t_\mathrm{R} + \tau_\mathrm{R} \cos k_\mathrm{R} t_\mathrm{R}}{\tau_\mathrm{cl}\cos k_\mathrm{R} t_\mathrm{R} - \tau_\mathrm{R} \sin k_\mathrm{R} t_\mathrm{R}} \right) \right] e^{-2\kappa_\mathrm{L} t_\mathrm{L}}.
\end{split}\label{Eq:Cladding:DispersionRelation}
\end{equation}
The solutions of Eq.~(\ref{Eq:Cladding:DispersionRelation}) are plotted in Fig.~\ref{Fig:EffectiveModeIndex}(a). For a certain value of $n_\mathrm{L}$, we can see that the effective mode index is low for small core thickness and increases with the core's thickness until it reaches the index of refraction of the core. Guided modes cease to exist for higher core thickness. In Fig.~\ref{Fig:EffectiveModeIndex}(b), we have replotted the effective mode index as a function of the core thickness for a waveguide with the same parameters as before and with $n_\mathrm{L} = \sqrt{0.4}$ [corresponding to the vertical line in Fig.~\ref{Fig:EffectiveModeIndex}(a)]. As expected, $n_\mathrm{eff}$ lies within the upper and lower bounds set by the refractive index of the cladding ($n_\mathrm{cl} = 1.0$) and core ($n_\mathrm{R} = 1.5$) materials. The dispersion relation for this mode is quite different from that of a mode in a traditional photonic waveguide. We clearly see that the effective mode index rises steeply with the thickness of the core layer and reaches the value of the core's index of refraction at $t_\mathrm{R} \approx 0.065\lambda$. The high effective mode index at this point signifies extremely good confinement of light to the core. In the simulation presented here, the thickness of the \lhm{} cladding layer is $0.033\lambda$, so that the total thickness of the waveguide structure does not exceed $\lambda/10$. This result confirms the existence of a well-confined guided mode with high effective mode index in a waveguide with subwavelength dimensions.

The addition of a thin layer of left-handed material to the cladding of a dielectric photonic waveguide thus opens a new way to miniaturization of photonic integrated circuits, advancing their integration with nanophotonic, plasmonic and electronic device elements. With electromagnetic energy largely confined to the right-handed core, which can be made from a dielectric with small losses, the proposed structure does not suffer from the inherent losses that seriously limit the propagation distance in plasmonic waveguides. If excited with a diffraction-limited beam through ATR, electromagnetic energy can be squeezed in the narrow waveguide due to a lensing effect,\cite{Foteinopoulou-2007} leading to very high intensities and enhanced nonlinear phenomena if a nonlinear material were present in the structure. The proposed waveguide could also be used as a very sensitive temperature or biophotonic sensor, since a small change in temperature or biological agent can increase the optical path length beyond the critical point in Fig.~\ref{Fig:EffectiveModeIndex}(b), effectively turning off the propagation.

This work was partially supported by BelSPO under
grant No.\ IAP-VI/10. P.\ T.\ is a Ph.D. Fellow of the FWO-Vlaanderen. The authors thank Guy Van der Sande for useful discussions and the anonymous Reviewer \#1 for providing some most interesting comments.



%
%
%
%

\end{document}